\documentclass{PoS}
\newcommand{\psib}{{\overline{\psi}}}
\usepackage{amssymb}
\usepackage{fontenc}
\usepackage{times}
\usepackage{mathptmx}
\usepackage{graphicx}

\title{Exotic Quantum Critical Points with Staggered Fermions}

\author{\speaker{Venkitesh Ayyar} \thanks{Work done in collaboration with Shailesh Chandrasekharan. The material presented here is based upon work supported by the U.S. Department of Energy, Office of Science, Nuclear Physics program under Award Number DE-FG02-05ER41368. This research was done using resources provided by the Open Science Grid, which is supported by the National Science Foundation and the U.S. Department of Energy's Office of Science.}\\
Duke University, Durham NC, USA.\\
E-mail: \email{vpa@phy.duke.edu}}

\abstract{We study two flavors of massless staggered fermions interacting via an on-site four-fermion interaction and argue that the model contains an exotic quantum critical point separating the perturbative massless phase from a massive fermion phase at strong couplings where the fermion bilinear condensate remains zero. We believe that no spontaneous symmetry breaking occurs at the transition. We have extensive calculations in three Euclidian dimensions that are consistent with the existence of a single second order phase transition separating the two phases. Although mean field theory suggests that this transition will turn first order at sufficiently large number of dimensions, preliminary results suggest that the transition remains second order in four-dimensions.}

\FullConference{The 33nd International Symposium on Lattice Field Theory,\\
		14-18 July, 2015\\
		Kobe International Conference Center, Kobe, Japan}

\ShortTitle{Exotic quantum critical points}

\begin{document}
\maketitle

\section{Introduction}
It is well known that chiral symmetries forbid the introduction of fermion mass terms. The conventional mode for fermion mass generation in the Standard Model involves the Spontaneous Symmetry Breaking(SSB) of chiral symmetries.
In the weak sector of the Standard Model, this is achieved by introducing a Mexican-hat potential. In the strong sector, mass generation is achieved by Dynamical Spontaneous Symmetry Breaking. 
In both cases, SSB is signalled through a non-zero fermion bilinear condensate. Our work aims to explore an alternate mechanism for fermion mass generation without SSB.

Four-fermion field theories in three dimensions are known to have exotic quantum critical points \cite{warr,hands}. Previous studies of four-fermion field theories in three dimensions have shown the existence of an interesting phase structure with zero condensates in the strong coupling limit \cite{anna}. 
We study such a model of staggered fermions in 3 Euclidean dimensions and find a single 2nd order transition separating the massless and massive phases. A similar result was found
in an extended Hubbard model on a honeycomb lattice \cite{cenke}.

\section{Model}
We study a simple four-fermion model containing two flavors of staggered fermions whose Euclidean action is given by
\begin{equation}
S = \sum_{i=1,2}\ \sum_{x,y} {\psib}_{x,i} \ M_{x,y} \ \psi_{y,i}\ - U \ \sum_{x} \Big\{ \psib_{x,1}\psi_{x,1}\psib_{x,2}\psi_{x,2}\Big\} 
\label{act}
\end{equation}
where $\psib_{x,i}, \psi_{x,i}, i = 1,2$ are four independent Grassmann valued fields. The matrix $M$ is the well known staggered fermion matrix given by
\begin{equation}
M_{x,y} \ =\  \sum_{\hat{\alpha}} \frac{\eta_{x,{\hat{\alpha}}}}{2}\ [\delta_{x,y+\hat{\alpha}} - \delta_{x,y-\hat{\alpha}}]
\label{staggered}
\end{equation}
where $x \equiv (x_1,x_2,x_3)$ denotes a lattice site on a $3$ dimensional cubic lattice and $\hat{\alpha} = \hat{1},\hat{2},\hat{3}$  represent unit lattice vectors in the three directions. The staggered fermion phase factors are defined as usual: $\eta_{x,\hat{1}}=1,, \eta_{x,\hat{2}}=(-1)^{x_1}$, and $\eta_{x,\hat{3}}= (-1)^{x_1+x_2}$. We will study cubical lattices of equal size $L$ in each direction with anti-periodic boundary conditions.

It  can be shown that, in addition to the usual space-time lattice transformations (translation, axis reversal and rotation) \cite{sym1,sym2}, the action is symmetric under internal $SU(4)$ transformations \cite{prev_proc}.

In this work, the observables we wish to measure are the correlators :
\label{corrdefs}
\begin{eqnarray}
C_a(x) &=& \langle \overline{\psi}_{0,1} \psi_{0,1} \overline{\psi}_{x_{odd},1} \psi_{x_{odd},1} \rangle \nonumber \\
C_b(x) &=& \langle \overline{\psi}_{0,1} \psi_{0,1} \overline{\psi}_{x_{even},2} \psi_{x_{even},2} \rangle \nonumber \\
F(x,y) &=& \langle \overline{\psi}_{x,1} \psi_{y,1} \rangle \label{corrs}
\end{eqnarray}

where expectation values are defined as
\begin{equation}
\Big\langle {\cal O} \Big\rangle = \frac{1}{Z}
\int [d\overline{\psi}\ d\psi]\ {\cal O}\ \mathrm{e}^{-S(\overline{\psi},\psi)}
\end{equation}
with $Z$ being the partition function. 

The $ SU(4) $ symmetry can be used to show that all other fermion bilinear correlators can be expressed in terms of $ C_a $ and $ C_b $. 
From the bosonic correlators defined in \ref{corrs}, we define the correlator ratios $ C_1 $ and $C_2 $ as follows:
\begin{eqnarray}
C_1 &=& \frac{C_a(L/2+1)}{C_a(1)} \nonumber \\
C_2 &=& \frac{C_b(L/2)}{C_b(0)} \label{obs_corrs}
\end{eqnarray}
A decay of the correlators would signal a zero condensate.

We also measure the average monomer density defined by :
\begin{equation}
\rho_m = \frac{1}{L^3} \ \sum_x\ \langle \overline{\psi}_{x,1}\psi_{x,1}\overline{\psi}_{x,2}\psi_{x,2} \rangle.
\end{equation}

\section{Computational Approach}
\label{sec4}

The traditional method for studying four-fermion field theories is by introducing an auxiliary field to convert the four-fermion term into a fermion bilinear. 
In this work we use an alternate approach called the Fermion bag approach \cite{fb1}.

\begin{figure}
\centering
\parbox{7cm}{
\includegraphics[width=\linewidth]{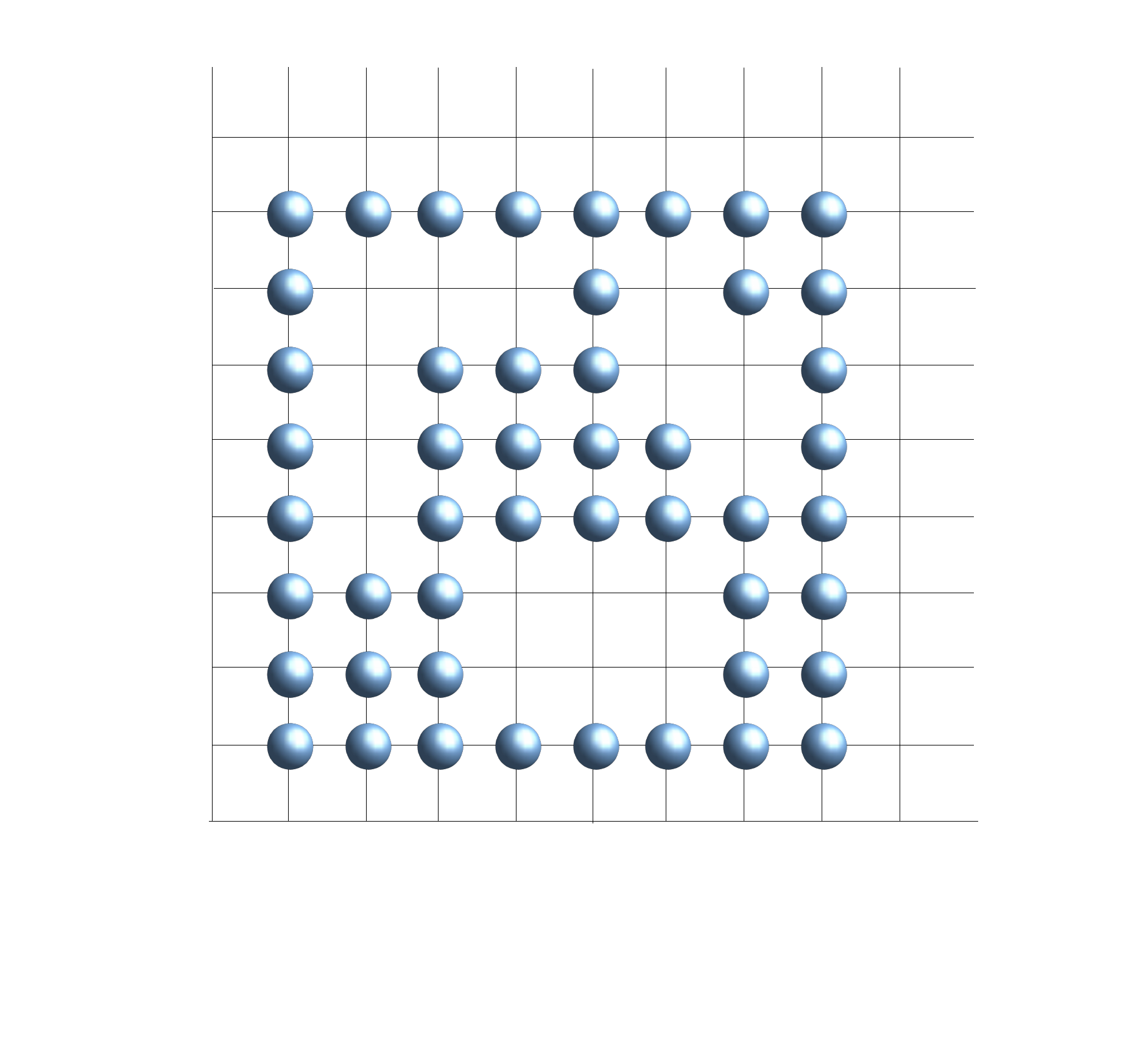}
\caption{\label{monoconf} An example of a monomer configuration $[n]$ showing free fermion bags on a two dimensional lattice. The blue circles denote the monomer sites.}}
\qquad
\begin{minipage}{7cm}
\includegraphics[width=\linewidth]{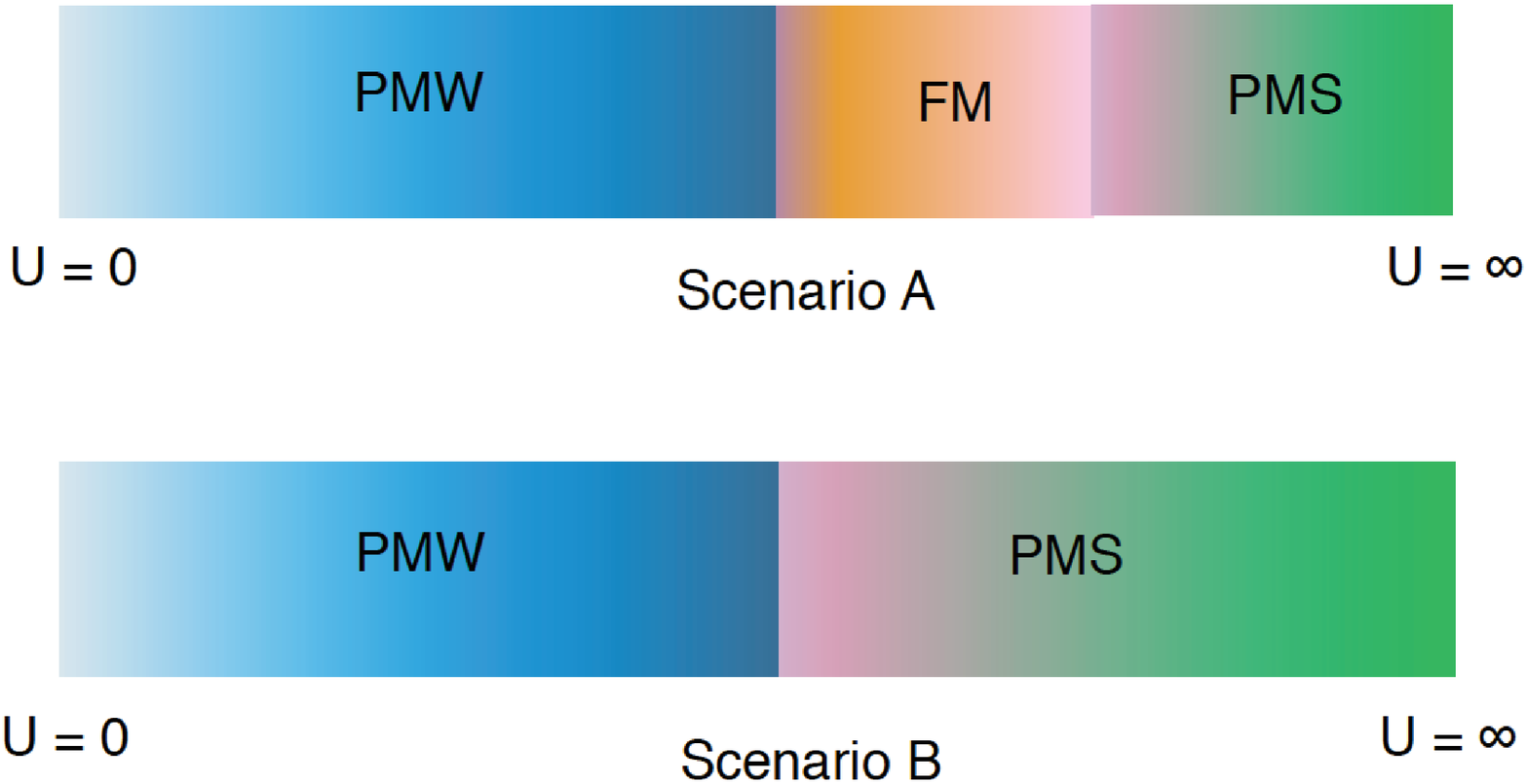}
\caption{\label{pdiag} The two possible phase diagrams for our model based on previous studies. Our work is consistent with scenario B with a second order transition between the PMW phase and the PMS phase.}
\end{minipage}
\end{figure}

In this approach, by defining a binary lattice field $n_x = 0,1$, we divide the lattice sites into monomers ($n_x=1$)  and free sites ($n_x=0$). 
Calling each such distribution a configuration $[n]$,  (Figure \ref{monoconf}) we can express our observables as a sum over these configurations. 
In other words, each configuration in the Markov Chain has a configuration weight and a value for the observables.
It can be shown that, the configuration weight and the observables can be expressed as a function of matrix determinants.  

In our previous work in \cite{our_paper}, we computed the configuration weights at each step and used this to compute the observables. As a result, whenever we retraced our path in the Markov chain
to arrive at the same configuration, the determinant computation had to be repeated, thus leading to a waste of computer time. We developed a new computational method to overcome this. 
In this new method, starting with a background configuration, we compute the configuration weights of all single perturbations that can be performed and store them. It can be shown that 
the weights of all subsequent configurations can be obtained from these. Once the perturbations are significantly big, we repeat this procedure.
This method enabled us to perform calculations on lattices upto size $ 60\times60\times60 $. However, the need to store the propagators results in larger memory requirements.

\section{Analysis and Results}

Using the Fermion bag approach, it is easy to visualize the different phases of the model in the strong and weak coupling limits. 
In 3D, since the four-fermion coupling is irrelevant, we must have a massless phase at weak couplings.
As explained in \cite{prev_proc}, all correlators must decay exponentially at strong couplings. This implies a massive phase where all bilinear condensates are zero.

Such exotic mechanisms of fermion mass generation have appeared in literature in the context of Yukawa models. Mean field calculations have shown a phase transition from a massless phase ( referred to as PMW or weak paramagnetic phase ) to an intermediate massive phase with a fermion bilinear chiral condensate ( referred to as FM phase ) and another transition from this intermediate phase to a new massive phase with zero fermion bilinear condensates ( referred to as PMS phase )\cite{p2}. 
Interestingly, other mean field calculations \cite{p3,p4} for 2 flavors in 3D give a single first-order phase transtion from the PMW to the PMS phases.
These two scenarios are shown in Fig \ref{pdiag}. In our work, we find a single second-order transition from the  PMW to PMS phases.

\begin{figure}[!htb]
\centering
\parbox{7cm}{
\includegraphics[width=\linewidth]{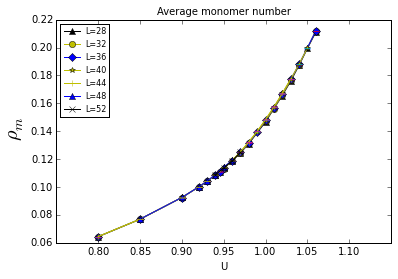}
\caption{\label{rhomono} Variation of the average monomer density $ \rho_m $ as a function of coupling $ U $ for lattices of size 28,32,36,40,44,48,52 .}}
\qquad
\begin{minipage}{7cm}
\includegraphics[width=\linewidth]{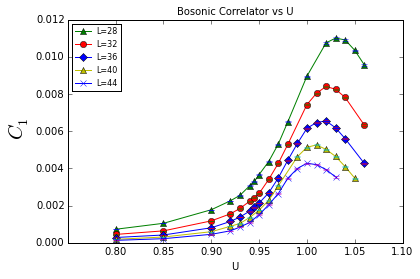}
\caption{\label{CvsU} Variation of the observable $ C_1 $ as a function of coupling $ U $.}
\end{minipage}
\end{figure}

\begin{figure}[!htb]
\centering
\parbox{7.5cm}{
\includegraphics[width=\linewidth]{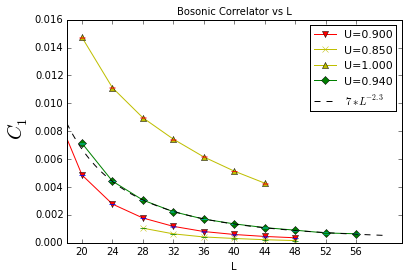}
\caption{\label{CvsL} Variation of the correlator ratio $ C_1 $ as a function of lattice size $ L $. Note the power law decay of the correlator ratio for $ U  \sim 0.945 $.}}
\qquad
\begin{minipage}{7.5cm}
\includegraphics[width=\linewidth]{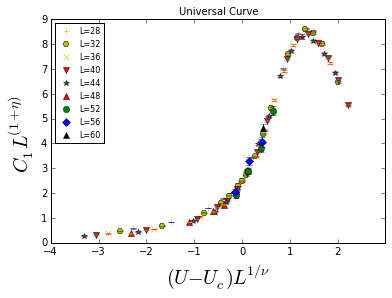}
\caption{\label{trans_order} Evidence for a second order phase transition}
\end{minipage}
\end{figure}

Looking at the behavior of the four-point condensate $ \rho_m $ as a function of the coupling $ U $ in Figure \ref{rhomono} it is clear that it is a smooth function, increasing from 0 for small couplings, rising sharply near $ U \sim 1 $ and approaching 1 for large couplings. The smooth nature of this curve is a hint that we have a single phase transition.
Figure \ref{CvsU} shows the behavior of the correlator ratio $ C_1 $ as a function of the coupling $ U $ for various values of the lattice size $ L$ . It is a smooth function of $ U $, reaching a maximum for $ U \sim 1 $. 
Figure \ref{CvsL} shows the behavior of $ C_1 $ as a function of $ L $ for different values of  $ U $. It can be seen that the correlator ratio decays with lattice size. For small couplings upto $ U \sim 0.95 $, 
the decay is a power law, implying a massless phase. Beyond $ U=0.96$, the decay in exponential, indicating the onset of the massive phase.

The correlator ratio $ C_2 $ shows a similar behavior.
All this points to a single phase-transition in the region close to $ U_c = 0.95 $.

From the theory of second-order phase transitions, near a second-order critical point, we know that correlators should vary continuously i.e.
\begin{equation}
C_1 \sim {L^{1+\eta}} f\left( (U-U_c)L^{\frac{1}{\nu}} \right) 
\end{equation}
Hence, a plot of $ \frac{C_1}{L^{1+\eta}} $ vs $ (U-U_c)L^{\frac{1}{\nu}} $ should be a smooth function.
Figure \ref{trans_order} shows that this is indeed true. This is clear evidence that we have a second-order phase transition between the PMW phase and the PMS phase.

Due to complications involving this 3 parameter fit, we have not been able to constrain the critical exponents effectively. Our preliminary estimate of the critical exponents are : 
$$ \eta = 1.0(1), \ \nu=1.2(1), \ U_c = 0.946(5) $$

\begin{figure}
\centering
\parbox{7cm}{
\includegraphics[width=\linewidth]{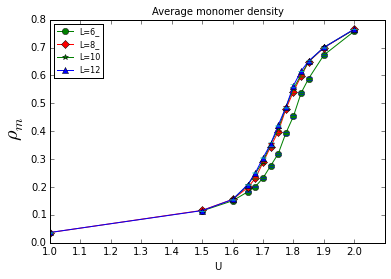}
\caption{\label{4d} Variation of the average monomer density $ \rho_m $ as a function of coupling $ U $ for lattices of size 6,8,10,12. It rises sharply near $ U=1.75$.}}
\end{figure}

We have also extended this analysis to 4D. We have results on lattices upto size $ 12^4 $. Preliminary results indicate a similar behavior for all observables.
The behavior of the four-point condensate $ \rho_m $ as a function of coupling is shown in Figure \ref{4d}. Its smooth nature hints at a single transition in 4D. 
However, more precise calcualtions are needed in 4D, to establish the phase structure.

\section{ Conclusions and Future work}
Our study of a lattice model with an on-site four-fermion interaction in 3 Euclidean dimensions shows that fermions can acquire a mass without the formation of fermion bilinear condensates.
There is no evidence of any Spontaneous Symmetry Breaking. The transition from massless to massive phase is confirmed to be second order.
Preliminary results in 4 dimensions sugggests a similar behavior. Further work on much larger lattices is required to confirm this. If confirmed in 4 dimensions, this would be a new 
mechanism for fermion mass generation without Spontaneous Symmetry Breaking that could be applicable in particle physics.

\end{document}